\documentclass[a4paper]{article}

\usepackage[english]{babel}
\usepackage[colorinlistoftodos, color=green!40, prependcaption]{todonotes}
\usepackage{amsthm}
\usepackage{mathtools}
\usepackage{physics}
\usepackage{xcolor}
\usepackage{graphicx}
\usepackage[left=23mm,right=13mm,top=35mm,columnsep=15pt]{geometry} 
\usepackage{adjustbox}
\usepackage{placeins}
\usepackage[T1]{fontenc}
\usepackage{lipsum}
\usepackage{csquotes}

\usepackage[utf8]{inputenc} 
\usepackage{newunicodechar}
\newunicodechar{​}{}
\DeclareUnicodeCharacter{2009}{\,}

\usepackage{hyperref}
\hypersetup{colorlinks,allcolors=black}

\usepackage{subcaption}

\usepackage{ragged2e}
\usepackage{sidecap}
\usepackage{amsmath}
\usepackage{bigints}
\usepackage[T1]{fontenc}
\usepackage[utf8]{inputenc}
\usepackage{authblk}

\DeclareUnicodeCharacter{2212}{\ensuremath{-}}

\begin{document}

\title{\huge Photo-Thermally Tunable Photon-Pair Generation in Dielectric Metasurfaces}
 
\author[1]{Omer Can Karaman\thanks{Equal contributions}}
\author[2,3]{Hua Li\thanks{Equal contributions}}
\author[1]{Elif Nur Dayi}
\author[2]{Christophe Galland\thanks{Correspondence email address:~chris.galland@epfl.ch}}
\author[1]{Giulia Tagliabue\thanks{Correspondence email address:~giulia.tagliabue@epfl.ch}}
\affil[1]{Laboratory of Nanoscience for Energy Technologies (LNET), STI, École Polytechnique Fédérale de Lausanne, 1015 Lausanne, Switzerland}
\affil[2]{Institute of Physics and Center for Quantum Science and Engineering, École Polytechnique Fédérale de Lausanne, 1015 Lausanne, Switzerland}
\affil[3]{State Key Laboratory of Coordination Chemistry, Key Laboratory of Mesoscopic Chemistry of MOE, School of Chemistry and Chemical Engineering, Nanjing University, 210023 Nanjing, China}

\maketitle

\begin{abstract}
Photon-pair sources based on spontaneous four-wave mixing (SFWM) in integrated photonics are often spectrally static. We demonstrate and model a fundamental thermo-optical mechanism that modulates photon-pair generation in amorphous silicon (a-Si) thin films and metasurfaces via SFWM. Femtosecond-pulsed excitation yields $g^{(2)}(0) > 400$ in unpatterned a-Si, confirming high-purity nonclassical emission. Resonant a-Si metasurfaces produce photon pairs at rates exceeding $3.8$~kHz under 0.6~mW pump power through Mie-type modes. Pump absorption induces localized heating that redshifts resonances, altering modal overlap and SFWM efficiency, leading to deviations from the quadratic power scaling expected in the undepleted regime. Coupled electromagnetic and heat-transfer simulations quantitatively reproduce these trends. Polarization-resolved measurements show nearly isotropic nonlinear responses, with $|\chi^{(3)}_{\mathrm{a\text{-}Si}}| \approx 3\times|\chi^{(3)}_{\mathrm{poly\text{-}Si}}|$. This work positions a-Si as a bright, CMOS-compatible quantum photonics platform and identifies thermo-optical detuning as a key mechanism that should be considered—and potentially harnessed—in integrated photon-pair sources.
\end{abstract}

\section{Introduction} \label{sec:Intro}

Entangled photon-pairs are foundational resources for quantum information technologies, enabling quantum communication, sensing, and computation protocols that surpass classical limits~\cite{kimble_quantum_2008,pan_multiphoton_2012,bouwmeester_experimental_1997}. Integrated sources of correlated photon-pairs, particularly those based on spontaneous four-wave mixing (SFWM), offer scalable and CMOS-compatible platforms for on-chip quantum photonics~\cite{elshaari_-chip_2017,son_generation_2025,sharapova_nonlinear_2023}. In this context, silicon-based materials have emerged as promising nonlinear media due to their high refractive index, mature fabrication infrastructure, and third-order nonlinearity \( \chi^{(3)} \)~\cite{foster_broad-band_2006,sharping_generation_2006,clemmen_continuous_2009}.

In particular, amorphous silicon (a-Si) has emerged as a compelling platform for nonlinear photonics, offering strong third-order optical nonlinearity~\cite{kuyken_nonlinear_2011,grillet_amorphous_2012}, strong resonant field confinement thanks to high dielectric constant \cite{kruk_nonlinear_2019,kuznetsov_optically_2016}, and compatibility with standard fabrication processes~\cite{leuthold_nonlinear_2010, bogaerts_silicon_2012}. Recent studies have leveraged these advantages to demonstrate low-threshold all-optical switching~\cite{leuthold_nonlinear_2010}, multi-wavelength metasurfaces~\cite{reineke_silicon_2019,arbabi_multiwavelength_2016}, and nonlinear imaging~\cite {schlickriede_nonlinear_2020}. However, photon-pair generation from a-Si metasurfaces remains underexplored. Unlike crystalline silicon, a-Si lacks long-range order, which relaxes symmetry constraints on its nonlinear tensor elements and can enable richer polarization dynamics and stronger local-field interactions due to its higher dielectric constant.
\begin{figure} [t!]
    \centering
  
   \includegraphics[]
   {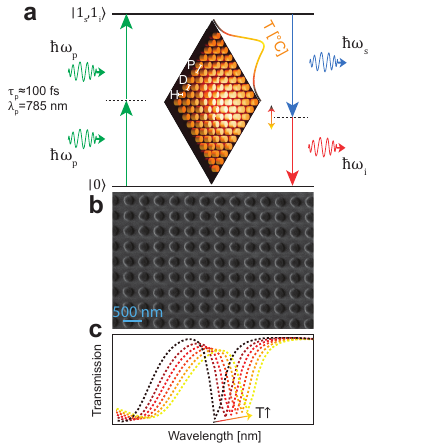}  \caption{\justifying{\textbf{Thermally tunable photon-pair generation in nonlinear a-Si metasurfaces.} 
\textbf{a} Illustration of degenerate SFWM in an a-Si metasurface on fused silica. ($H$, $D$, and $P$ denote the height, diameter, and periodicity of the disks, respectively). The pump-induced thermal shift alters resonance conditions and modal overlap, dynamically reconfiguring photon pair-emission. 
\textbf{b} Scanning electron microscope (SEM) image of a fabricated metasurface (scale bar: 500~nm).
\textbf{c} Conceptual depiction of the photo-thermo-optical resonance redshift.
 }}
    \label{fig1}
\end{figure}
In parallel, resonant dielectric metasurfaces based on Mie-type modes have shown significant promise for enhancing nonlinear optical processes~\cite{shcherbakov_photon_2019,sharapova_nonlinear_2023,lewi_thermal_2019}. These metasurfaces consist of Mie resonant nanostructures, which confine light to subwavelength volumes and exhibit sharp magnetic dipole (MD) and electric dipole (ED) resonances, thereby boosting nonlinear efficiencies via local field enhancement and offering new geometries for phase matching~\cite{santiago-cruz_photon_2021,zograf_stimulated_2020}. Integrating a-Si with such metasurfaces introduces a unique opportunity to simultaneously harness optical and structural nonlinearities arising from resonantly confined electromagnetic fields within the nanostructured geometry.

Moreover, the nonlinear optical response in such resonant structures is inherently sensitive to temperature due to the thermo-optical effect in silicon. Under resonant excitation, pump absorption leads to localized heating, which redshifts the resonance wavelength, as schematically illustrated in \textbf{Figure \ref{fig1}c}. This \emph{thermo-optical detuning} dynamically modifies the spectral overlap between the pump, signal, and idler modes, and thus the nonlinear overlap integral governing SFWM efficiency. While this feedback mechanism has been extensively studied in classical metasurface responses~\cite{karaman_decoupling_nodate,duh_giant_2020,tang_mie-enhanced_2021-1}, its direct impact on photon-pair generation has not been experimentally quantified. Importantly, this effect is not limited to silicon: any high-index material with appreciable pump absorption, such as transition metal dichalcogenides~\cite{chakraborty_advances_2019} or perovskites~\cite{wang_perovskite_2018}, will exhibit similar detuning-driven modifications to quantum emission. Even for sub-bandgap excitation, two-photon absorption can also lead to thermo-optical shifts \cite{prosad_single-_2023}. Understanding and modeling this thermo-optical coupling is therefore essential for the design of stable and efficient integrated photon-pair sources.


In this work, we investigate photon-pair generation by SFWM in a-Si and polycrystalline silicon (poly-Si) thin films, as well as in metasurfaces composed of a-Si nanodisks, using a femtosecond-pulsed laser (785\,nm center wavelength, $\approx$100\,fs pulse duration, \textbf{Figure \ref{fig1}a}). We observe second-order correlation values as high as \( g^{(2)}(0) > 400 \) from an unpatterned 100\,nm-thick a-Si film at 0.6~mW pump power, confirming low-noise nonclassical pair emission with negligible multi-pair contributions. At 4.8~mW pump power, poly-Si thin film exhibits higher \( g^{(2)}(0)\) than a-Si ($g^{(2)}(0)_{poly-Si}\approx160$ while $g^{(2)}(0)_{a-Si}\approx70$ ) but with significantly reduced brightness, linked to its narrower Raman scattering linewidth and weaker $\chi^{(3)}$. Resonant a-Si metasurfaces enable much higher photon-pair generation rates, inferred to exceed 3.8\,kHz at 0.6\,mW pump power, driven by ED and MD resonances that enhance modal overlap. We further demonstrate that photo-induced thermal shifts in resonance dynamically modulate the photon-pair generation efficiency, enabling thermally reconfigurable quantum emission. These effects are modeled using full-field electromagnetic simulations and a quasi-continuous-wave (CW) heat transfer model incorporating the temperature-dependent refractive index. Finally, polarization-resolved measurements reveal the nearly isotropic nonlinear response of a-Si, in contrast to the anisotropic behavior of poly-Si, allowing us to extract an effective nonlinear enhancement of \( |\chi^{(3)}_{\mathrm{a\text{-}Si}}| \approx 3 \times |\chi^{(3)}_{\mathrm{poly\text{-}Si}}| \). These findings establish a-Si as a high-efficiency photon-pair source and a tunable nonlinear material platform for reconfigurable quantum photonic devices.

\section{Results and Discussion} \label{sec:Results}

\begin{figure}[t!]
    \centering
  
   \includegraphics[]
   {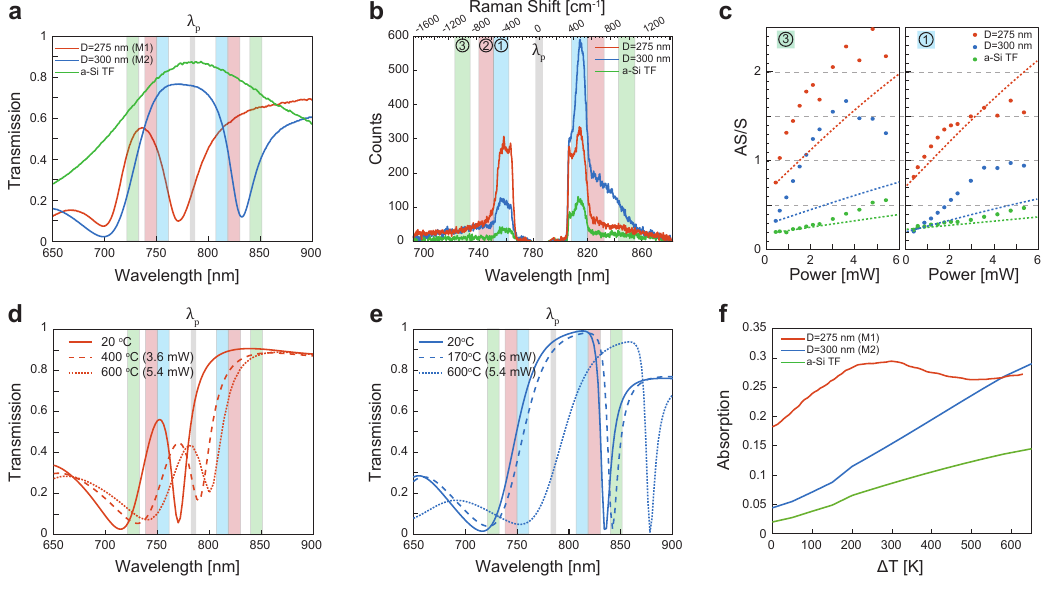}
    \caption{\justifying{\textbf{Elastic and inelastic optical response of a-Si thin films and metasurfaces.} 
\textbf{a} Measured transmission spectra for metasurface M1 (orange, $D = 275$~nm and $P = 380$~nm), M2 (blue, $D = 300$~nm and $P = 380$~nm), and a-Si thin film (green), all with $H = 100$~nm. 
 Shaded regions indicate the detection bands used for AS/S and photon-pair measurements (band 1: 750–763 / 808–820~nm; band 2: 740–751 / 822–835~nm; band 3: 722–733 / 844–854~nm). Gray region marks the pump spectrum at 785~nm. 
 \textbf{b} Raman spectra acquired in transmission at 0.4~mW pump power (5 pJ per pulse). The two objectives have a numerical aperture of 0.8, corresponding to a diffraction-limited spot diameter of $\sim1.2~\mu$m. 
\textbf{c} Power-dependent anti-Stokes to Stokes intensity ratios (AS/S) for M1, M2, and a-Si thin film in bands 1 (right) and 3 (left). Dashed lines correspond to predictions using the Boltzmann relation $I_{\mathrm{AS}}/I_{\mathrm{S}} \propto F\exp(-\hbar\Omega / k_B T)$ (neglecting thermo-optical effects), scaled by a constant factor $F$ for each case to fit at low powers. Power was converted to effective temperature using Eq.~(\ref{eq1}) and room-temperature absorption values: $Abs_{M1} = 0.18$, $Abs_{M2} = 0.05$, $Abs_{a-Si\ TF} = 0.03$.
\textbf{d,e} Simulated transmission spectra of M1 and M2 at different temperatures (solid: 25$^\circ$C; dashed: intermediate; dotted: 600$^\circ$C), showing resonance redshift. 
\textbf{f} Simulated absorption at 785~nm versus temperature rise $\Delta T$ for M1, M2, and a-Si thin film.}}
    \label{fig2}
\end{figure}

 We first characterize the power-dependent elastic (transmission) and inelastic (Raman) responses of an unpatterned 100~nm a-Si thin film and resonant metasurfaces (SEM in \textbf{Figure~\ref{fig1}b}). \textbf{Figure~\ref{fig2}a} shows transmission spectra for metasurfaces consisting of a-Si nanodisks with diameters $D=275$~nm (M1, orange) and 300~nm (M2, blue), periodicity $P=380$~nm, and the a-Si thin film (green). M1 and M2 display MD and ED resonances; ED wavelengths increase with disk size, while MD modes occur at shorter wavelengths set by geometry and material \cite{park_structural_2017}. Specifically, M1 supports an MD resonance at $\sim$700~nm and an ED resonance at $\sim$770~nm, and M2 supports an MD resonance at $\sim$700~nm and an ED resonance at $\sim$835~nm (field profiles in \textbf{SI Note~1}). The thin film lacks discrete modes but shows a broad $\sim$780~nm peak in transmission from constructive thin-film interference (\(2nH \approx \lambda\), $n\approx4$, $H=100$~nm), which, though non-localized, can influence pump field enhancement and collection efficiencies of Raman or SFWM signals \cite{santos_entangled_2024}.

\textbf{Figure \ref{fig2}b} shows Raman spectra for the same three samples measured at a pump power of 0.4~mW at 785~nm. All structures display both Stokes (S) and anti-Stokes (AS) Raman sidebands, but the metasurfaces, particularly M1 and M2, exhibit significantly enhanced AS and S scattering near their MD and ED resonant modes, confirming the increased photonic density of states at these wavelengths. Three pairs of spectral windows symmetrically shifted in energy from the pump are experimentally selected with bandpass filters for coincidence measurements: detection band 1 - blue shaded area: 750-763~nm / 808-820~nm  (corresponding to the transverse optical (TO) phonon mode at 470~cm$^{-1}$); detection band 2 - red shaded area: 741-751~nm / 820-835~nm (near the second-order longitudinal acoustic (2LA) mode at 561~cm$^{-1}$); detection band 3 - green shaded area: 722-733~nm / 844-854~nm (corresponding to higher-order Raman sidebands near the second-order transverse optical (2TO) region) \cite{karaman_decoupling_nodate}. 

When the phonon population is in thermal equilibrium, the anti-Stokes to Stokes (AS/S) Raman intensity ratio will scale with the Boltzmann distribution, \( I_{\mathrm{AS}}/I_{\mathrm{S}} \propto \exp(-\hbar\Omega / k_B T) \)~\cite{hart_temperature_1970}, where \( \hbar\Omega \) is the phonon energy, \( k_{\mathrm{B}} \) is the Boltzmann constant, and \( T \) is the absolute temperature. The temperature rise \( \Delta T_i \) for a given optical pump power is estimated using an analytical heat balance model under quasi-steady-state conditions. Although the excitation source is a pulsed laser (785~nm, $\sim$100 fs), the high repetition rate (80~MHz) ensures quasi-continuous-wave (CW) heating behavior with negligible thermal relaxation between pulses. The steady-state temperature rise is given by~\cite{baffou_nanoscale_2010}:
\begin{equation} \label{eq1}
\Delta T_i = \frac{\sigma_{\mathrm{abs}} I}{4\pi R_{\mathrm{eq}} \beta \kappa} = \frac{\text{(absorptance)} \cdot P}{4\pi R_{\mathrm{eq}} \beta \kappa},
\end{equation}
where \( P \) is the pump power, \( \kappa \) is the thermal conductivity, \( \beta \) is a geometric thermal factor, and \( R_{\mathrm{eq}} \) is the effective radius of the structure. The denominator can be interpreted as an equivalent thermal capacitance \( C_{\mathrm{eq}} \), calibrated using measured absorptance values and experimentally inferred damage thresholds.

In \textbf{Figure~\ref{fig2}c}, we compare the experimentally measured AS/S ratios with the expected thermal equilibrium scaling (dashed lines) under the two following `static' assumptions. First, the temperature vs. pump power is predicted using Equation~\ref{eq1} considering \textit{constant} room-temperature absorption coefficients ($Abs_{M1} = 0.18$, $Abs_{M2} = 0.05$, $Abs_{a-Si\ TF} = 0.03$). Second, the resulting Boltzmann curves are scaled using a \textit{constant} factor \( F \) to match the low-power data. This prefactor accounts for the different optical densities of states and setup detection efficiency at the S and AS wavelengths, in the low power limit. We find that the experimental AS/S ratios exhibit a much steeper and nonlinear increase with pump power than this naive model predicts, particularly for the resonant metasurfaces. 
This discrepancy is understood as a manifestation of the dynamical shift of resonant modes due to thermo-optical effects. It modifies both the absorption coefficient at the pump wavelength and the field enhancement at the Stokes and anti-Stokes wavelengths as a function of pump power, making the overall behavior highly nonlinear.

This interpretation is directly supported by the simulated temperature-dependent transmission spectra of M1 and M2 shown in \textbf{Figure~\ref{fig2}d} and \textbf{\ref{fig2}e}. 
As the temperature increases from 25 $^\circ$C to 600 $^\circ$C, both metasurfaces exhibit a progressive redshift in their resonance wavelengths due to the thermo-optic effect of a-Si (\textbf{Figure~\ref{fig2}d,e}). For M1 (\textbf{Figure~\ref{fig2}d}), the MD resonance lies near the anti-Stokes detection bands (bands 1 and 3) at room temperature and redshifts further into these bands with increasing power, thereby enhancing anti-Stokes scattering. The ED resonance, centered near 770~nm, contributes to a power-dependent increase in absorption, as verified in \textbf{Figure~\ref{fig2}f}, causing the temperature to rise more rapidly than predicted by the static model. Together, these two effects drive a steep rise in the AS/S ratio and lead to strong deviations from Boltzmann scaling. 
The dynamical behavior of M2 can similarly be traced back to the power-dependent resonance shifts computed in  \textbf{Figure~\ref{fig2}e}. 
These results confirm that the power-dependent AS/S ratio in both metasurfaces is governed by the evolving spectral overlap between the resonances and the Raman emission bands -- a mechanism that also drives the nonlinear power dependence of photon-pair generation, as we show below.
Another factor contributing to the breakdown of Boltzmann scaling in \textbf{Figure~\ref{fig2}c} is the contribution of SFWM photon-pairs to the collected spectrum. These photons always come in pairs, and their AS/S ratio does not obey Boltzmann statistics. Indeed, at low pump powers, where heating is modest, we find that the measured AS/S ratio already departs from the model prediction. 
Consequently, the observed AS/S ratio represents a hybrid signature of both thermally activated Raman scattering and SFWM processes \cite{parra-murillo_stokes--anti-stokes_2016}.

\begin{figure} [t!]
    \centering
  
   \includegraphics[]
   {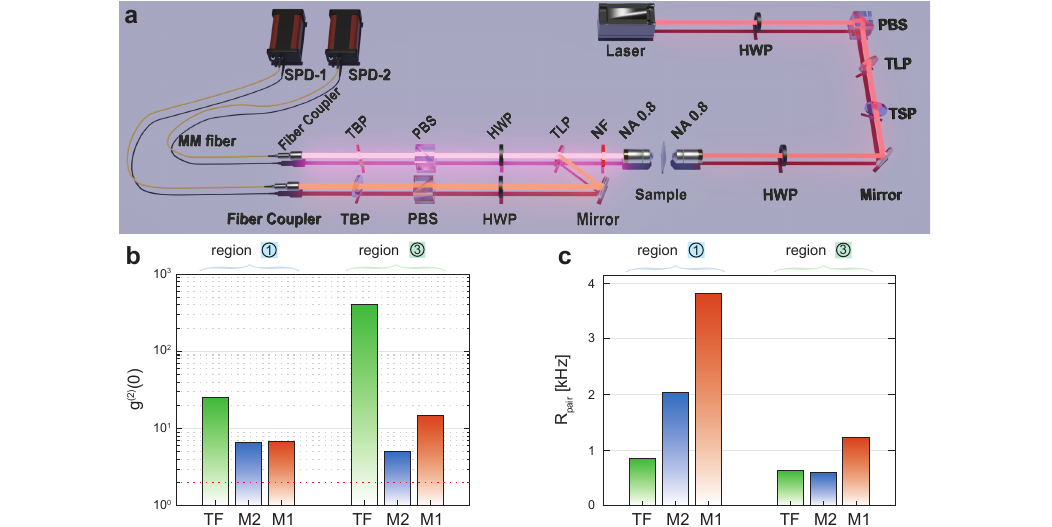}
    \caption{\justifying{\textbf{Coincidence measurement scheme and photon-pair generation benchmarks.} 
\textbf{a} Experimental setup for coincidence measurements (see \textbf{Methods} for the list of optical elements). 
\textbf{b} Second-order correlation \( g^{(2)}(0) \) for photon-pairs from a-Si thin film (green), M1 (orange), and M2 (blue) at detection bands 1 (left) and 3 (right), measured at 0.6~mW vertically polarized pump. The red dashed line indicates the classical limit \( g^{(2)}(0) = 2 \), under the assumption that individual channels feature thermal statistics \cite{velez_preparation_2019}.
\textbf{c} Photon-pair generation rates \( R_{\mathrm{pair}} \) under the same conditions as in \textbf{b}.}}
    \label{fig3}
\end{figure}

\begin{figure} [t!]
    \centering  
   \includegraphics[]
   {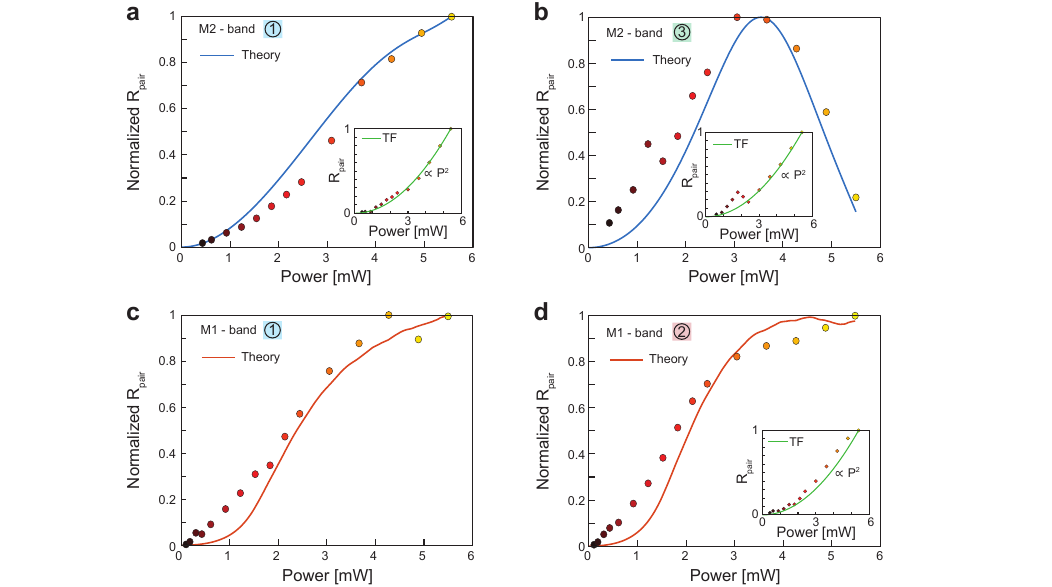}
    \caption{\justifying{\textbf{Thermo-optical effects on power-dependent photon-pair generation.} 
\textbf{a–d} Power-dependent \( R_{\mathrm{pair}} \) (full dots: measurements, normalized to their maximum values) and theoretical predictions (full lines) for: 
\textbf{a} M2 at detection band 1; 
\textbf{b} M2 at detection band 3; 
\textbf{c} M1 at detection band 1; 
\textbf{d} M1 at detection band 2. 
Insets are the corresponding thin film a-Si responses with quadratic scaling (dashed line).}}
    \label{fig4}
\end{figure}

We now characterize the non-classicality of the SFWM photon pair emission and measure the coincidence rate between Stokes and anti-Stokes channels with the setup shown in \textbf{Figure \ref{fig3}a} (See Methods for details).
All experiments were conducted using a 785~nm, 100~fs, 80~MHz pulsed laser as the excitation source. \textbf{Figure \ref{fig3}b} presents the zero-delay second-order correlation function, \( g^{(2)}(0) \) (equivalent to the CAR, coincidence-to-accidental ratio, for photon-pairs collected from the thin film, M1, and M2 metasurfaces, in detection bands 1 (left) and 3 (right). The a-Si thin film exhibits remarkably high \( g^{(2)}(0) \) values—reaching 400 in detection band 3—demonstrating high-purity pair generation. The metasurfaces show lower but still non-classical correlations, with \( g^{(2)}(0) \) ranging from approximately 5 to 20. 

\textbf{Figure \ref{fig3}c} shows the absolute photon-pair generation rates \( R_{\mathrm{pair}} \) measured at 0.6\,mW average pump power. It is further noted that \( R_{\mathrm{pair}} \) depends strongly on detector efficiency and the insertion losses associated with the filtering components. In detection band 1, M1 exhibits the highest pair generation rate, reaching 3.8\,kHz. This corresponds to a pair creation probability of $5\times10^{-5}$ per pulse, where multiple pair generation is negligible. 
The improved generation rate in M1 and M2 highlights the role of suitably tuned Mie-type modes in enhancing SFWM efficiency. We attribute the trade-off between brightness (high $R_{\mathrm{pair}}$) and purity (high $g^{(2)}(0)$) to the increase in uncorrelated Raman scattered photon emission from the metasurfaces, contributing to more accidental counts. 

\textbf{Figure~\ref{fig4}a--d} reveal how the photon-pair generation rate, $R_{\mathrm{pair}}$, evolves with pump power for different metasurface–detection band combinations. These measurements directly reveal thermo-optical tuning as the mechanism behind the observed power dependence of SFWM efficiency. In the absence of Mie-type resonances, the a-Si thin film exhibits a quadratic dependence of the photon-pair rate on pump power, consistent with undepleted SFWM (\textbf{Figure~\ref{fig4}a-d} insets show the measurements from a-Si thin film). In contrast, resonant metasurfaces deviate from this behavior once pump-induced thermo-optical tuning modifies the spectral overlap between the pump, signal, and idler modes. This spectral shift can enhance or suppress the photon-pair rate depending on whether it moves the resonances into or out of optimal spectral alignment with the signal and idler bands. Experimental measurements (dots) for the film are compared to a quadratic dependence \( R \propto P^2 \) expected from ideal undepleted SFWM. The metasurface data are compared to a model based on the mode overlap integral, which governs the quantum mechanical transition probability per unit time from the vacuum state to a two-photon final state, mediated by the nonlinear interaction Hamiltonian. The resulting expression for the emission rate into signal and idler modes is given by \cite{gerry_introductory_2004}:
\begin{equation}
\Gamma_{s,i} = \frac{2\pi}{\hbar} \left| \langle f | \hat{H}_{\mathrm{int}} | i \rangle \right|^2 \delta(\omega_s + \omega_i - 2\omega_p),
\end{equation}
where \( |i\rangle \) and \( |f\rangle \) denote the initial vacuum and final two-photon states, and the delta function enforces energy conservation between the pump and generated photons.

The term of the interaction Hamiltonian that is responsible for the SFWM process measured here can be approximated by: \cite{boyd_chapter_2008,quesada_beyond_2022}:
\begin{equation}
\hat{H}_{\mathrm{int}} = \frac{1}{2} \varepsilon_0 \int_V \chi^{(3)} : \mathbf{E}_p(\mathbf{r}) \mathbf{E}_p(\mathbf{r}) \hat{\mathbf{E}}_s^\dagger(\mathbf{r}) \hat{\mathbf{E}}_i^\dagger(\mathbf{r}) \, d^3\mathbf{r} + \text{H.c.}
\end{equation}
Here, \( \mathbf{E}_p(\mathbf{r}) \) is the classical pump field, \( \hat{\mathbf{E}}_s^\dagger \) and \( \hat{\mathbf{E}}_i^\dagger \) are the quantized signal and idler fields, and \( \chi^{(3)} \) is the third-order susceptibility, a fourth-rank tensor that must be contracted with the four vectorial quantities on its right (leading up to 81 terms). Under a mode decomposition framework, this transition amplitude reduces to a nonlinear overlap integral \cite{weissflog_tunable_2024,poddubny_generation_2016,clemmen_continuous_2009}:
\begin{equation}
\Gamma_{s,i} \propto \left| \int_V \chi^{(3)}(\mathbf{r}) \, \mathbf{E}_p(\mathbf{r}) \cdot \mathbf{E}_p(\mathbf{r}) \cdot \mathbf{E}_s^*(\mathbf{r}) \cdot \mathbf{E}_i^*(\mathbf{r}) \, d^3\mathbf{r} \right|^2.
\end{equation}

The spatial field distributions at the pump, signal, and idler wavelengths are computed using full-wave simulations in the COMSOL Wave Optics Module. To account for thermal effects, temperature-dependent refractive index variations are incorporated into the simulations, allowing dynamic recalculation of the field profiles and corresponding overlap integrals as a function of laser power. This enables quantitative modeling of both resonance-enhanced SFWM and its modification due to thermo-optical tuning.
Remarkably, this theoretical model reproduces well the deviations from quadratic power scaling in \textbf{Figure \ref{fig4}} (solid curves), and even captures the non-monotonic behavior in \textbf{Figure \ref{fig4}b} due to the redshift of M2’s ED resonance.

\textbf{Figure \ref{fig4}c} and \textbf{d} show the behavior of M1 in detection bands 1 and 2. In both cases, \( R_{\mathrm{pair}} \) begins to saturate around 3.6~mW (corresponding to $\Delta T \approx 380~^\circ$C). The overlap-integral-based theory closely matches these trends and explains the earlier saturation in detection band 2 as arising from the detuning of M1’s ED mode around the pump wavelength (see \textbf{Figure \ref{fig2}d}, a detailed breakdown of the temperature–overlap relationship for these cases is provided in \textbf{SI Note 2},). Due to the higher imaginary part of the thermo-optical coefficient of a-Si around 770~nm compared to 835~nm \cite{karaman_decoupling_2024}, the ED mode of M1 broadens and reduces the enhancement in \( R_{\mathrm{pair}} \) compared to M2 metasurface. 

Importantly, thermo-optical tuning is perfectly reversible in the regime investigated here, meaning that the system recovers its initial states after cooling back to room temperature without causing structural damage. Altogether, these results demonstrate that photon-pair generation in a-Si metasurfaces is not only enhanced by dipolar resonances but also dynamically tunable via thermo-optical effects.

We also compared polarization-resolved SFWM in 100\,nm-thick a-Si and poly-Si films at 4.8\,mW pump power. 
(see \textbf{SI Note 4}). Poly-Si exhibited the highest \( g^{(2)}(0) \) values (reaching up to 160) in certain co-polarized channels, whereas a-Si consistently produced higher photon-pair generation rates---up to 6.5\,kHz in VV under V-polarized pumping and more than nine times higher counts than poly-Si under the same condition (\textbf{SI Figure S4}). This reflects a material-dependent brightness/purity trade-off: the broader Raman spectrum of a-Si, arising from LA, LO, and TO phonon modes and their higher-order combinations, increases spectral overlap with the SFWM detection windows, boosting count rates but also accidental coincidences, thereby lowering \( g^{(2)}(0) \) (see \textbf{SI Figure S5}). In contrast, poly-Si’s narrower Raman peak reduces background photons and yields higher correlation purity, but with significantly reduced brightness \cite{ahn_fabrication_2011,vavrunkova_study_2010,hyopark_tempered_2015}. Note that each Raman-active phonon mode also contributes to a resonant term in the $\chi^{(3)}$ tensor, which has recently been leveraged for the production of polarization-entangled SFWM photon pairs in bulk diamond \cite{vento_how_2025,freitas_microscopic_2023}, opening more opportunities for silicon-based metasurfaces.

In both amorphous and poly-crystalline silicon, the polarization response is governed by the dominant \( \chi^{(3)}_{xxxx} \) and \( \chi^{(3)}_{xxyy} \) tensor elements. The poly-Si thin film behaves as an isotropic medium due to grain size much smaller than the wavelength (changes from 3 to 18~nm depending on the crystallographic planes, see \textbf{SI Note 3} for details). From dominant-channel brightness ratios, we estimate an effective nonlinearity enhancement of \(|\chi^{(3)}_{\mathrm{a\text{-}Si}}| \approx 3\times|\chi^{(3)}_{\mathrm{poly\text{-}Si}}|\).
These results underline the importance of material choice in balancing brightness and purity for integrated quantum sources: a-Si is advantageous for high-flux applications, while poly-Si is suited for scenarios requiring maximal non-classicality.


In summary, our results demonstrate that photon-pair generation in a-Si metasurfaces is strongly influenced by optically induced thermal shifts of the Mie resonances. Unpatterned 100\,nm a-Si films exhibit nonclassical emission with \( g^{(2)}(0) \) values up to 400, surpassing typical silicon microrings and waveguides~\cite{clemmen_continuous_2009,sharping_generation_2006,harada_generation_2008,guo_high_2017,matsuda_-chip_2014,harris_integrated_2014,heuck_unidirectional_2019} and underscoring the potential of ultrathin a-Si for nonlinear quantum optics. Patterning metasurfaces that feature ED and MD resonances further enhances SFWM efficiency, achieving photon-pair rates above 3.6\,kHz at 0.6\,mW pump power. Unlike the ideal quadratic scaling expected for SFWM, metasurfaces exhibit saturation and decline at elevated powers, a behavior well explained by dynamical photo-thermo-optic shifts of the resonances that alter mode overlap. This interpretation is supported by thermal modeling and temperature-dependent full-field simulations. Raman backgrounds further shape performance: the broad spectrum of a-Si introduces additional noise photons, while poly-Si’s narrower Raman lines allow higher \( g^{(2)}(0) \) despite reduced pair brightness (with effective \(|\chi^{(3)}_{\mathrm{a\text{-}Si}}| \approx 3\times|\chi^{(3)}_{\mathrm{poly\text{-}Si}}|\)). This establishes a fundamental brightness/purity trade-off. More broadly, our findings link photo-thermally driven optical dynamics---well known in classical studies~\cite{duh_giant_2020,heuck_unidirectional_2019,cotrufo_passive_2024,jiang_efficient_2024,naef_light-driven_2023,tsoulos_self-induced_2020,archetti_thermally_2022,karaman_ultrafast_2024}---to quantum light generation, revealing a route to compact, reprogrammable photon-pair sources with wavelength trimming, fabrication-tolerance compensation, and fast power-controlled switching between brightness and purity regimes, features absent in static integrated devices.

\section{Methods} \label{methods}
\textbf{Sample fabrication:} The metasurfaces were fabricated on 550\,µm thick fused silica substrates. Uniform 100\,nm a-Si and poly-Si films were deposited via plasma-enhanced chemical vapor deposition (PECVD) at 550\,$^\circ$C and 625\,$^\circ$C , respectively. This deposition method provided excellent thickness control and surface conformity, essential for achieving high-quality metasurface resonances. To define the nanodisk arrays, a 120\,nm layer of ZEP 520A resist was spin-coated and patterned using electron-beam lithography (EBPG5000+ system) operated at 100\,kV with a dose of 200\,\textmu C/cm$^2$. The developed resist pattern served as a hard mask for transferring the nanodisk layout into the a-Si layer.

Anisotropic etching was performed using an argon ion beam etcher (Veeco Nexus IBE350), operated at 170\,V and 175\,mA, to etch the exposed a-Si regions. To eliminate a-Si redeposition on the backside of the chips during the process, the rear surface was simultaneously etched under the same conditions. Following pattern transfer, the residual resist was stripped by immersion in acetone. A final surface cleaning step was carried out using low-power microwave plasma treatment in an oxygen-rich environment (Tepla 300, 500\,W, 400\,mL/min O$_2$) to remove any remaining organic contaminants without compromising the structural integrity of the nanodisks.
\\
\textbf{Experimental setup for coincidence measurements:} \label{setup}
The photon-pair coincidence measurements were performed using a free-space transmission setup centered around a mode-locked Ti:Sapphire laser (Coherent Mira 900), operating at 785\,nm with 100\,fs pulse duration and 80\,MHz repetition rate, which can be seen in \textbf{Figure \ref{fig3}a}. The laser beam was first passed through a half-wave plate (HWP) and a polarizing beam splitter (PBS) to control the power of the incident light. After spectral cleaning with a tunable shortpass (TSP) and longpass filter pair (TLP), the beam was focused onto the sample using a 100$\times$ objective lens (NA = 0.8). The sample was mounted between two opposing objectives (100$\times$/0.8 NA), allowing for the collection of the transmitted signal. Transmitted pump light was suppressed using a 785\,nm/33\,nm notch filter (NF) placed in the detection path. In addition, a tunable longpass filter (TLP) was used to remove the residual pump and select the Stokes and anti-Stokes sidebands. The collected light was then split into two detection arms, each equipped with a 10\,nm bandwidth tunable bandpass filter (TBP), allowing spectral selection of signal and idler photon wavelengths corresponding to specific SFWM emission regions. Three HWPs and two PBS cubes were added in the excitation and two collection path respectively for checking the polarization dependence of the photon-pairs. Each filtered channel was coupled into a single-mode (SM) fiber and directed to a silicon avalanche photodiode (SPD) (Excelitas SPCM-AQRH). Time delay of two channels were controlled using the ID900 time controller (ID Quantique). Coincidence histograms were recorded and processed to extract the photon correlation statistics. For the evaluation of  $g^{(2)}(0)$, the background noise was subtracted from the raw coincidence histograms. Background counts were estimated by averaging the coincidence counts in distant temporal windows, outside the central correlation peak, to account for ambient noise and detector dark counts. The resulting background-corrected histograms were then used to compute $g^{(2)}(0)$ as:
\begin{equation}
g^{(2)}(0) = \frac{C_{\text{bg-subtracted}}(0)}{C_{\text{acc}}},
\end{equation}
where $C_{\text{bg-subtracted}}(0)$ is the coincidence count at zero delay after background subtraction, and $C_{\text{acc}}$ is the average accidental coincidence count in uncorrelated side peaks, used as a normalization factor.

For determining the true coincidence rates $R_{\text{coincidence}}$, both background and accidental coincidences were subtracted. Accidental coincidences, primarily arising from uncorrelated photons within the detector timing window, were estimated from side peaks in the histogram and removed to isolate the photon-pair generation signal. The timing gate used for peak integration was 100~ps. 

The total collection efficiency for each photon detection path is denoted by \(\eta_{\text{collection}}\), and accounts for the cumulative losses across all optical components from the sample to the detector. We define it as:

\[
\eta_{\text{collection}} = \eta_{\text{obj}} \cdot \eta_{\text{fs}} \cdot \eta_{\text{fiber}} \cdot \eta_{\text{det}}
\]

\noindent where \(\eta_{\text{obj}}\) is the transmission efficiency through the microscope objective, \(\eta_{\text{fs}}\) represents the combined throughput of free-space optical elements (including mirrors, notch, dichroic, and bandpass filters), \(\eta_{\text{fiber}}\) is the efficiency of coupling into and transmitting through single-mode fibers, and \(\eta_{\text{det}}\) is the quantum efficiency of the SPDs. Based on manufacturer datasheets and experimental calibration, we estimate \(\eta_{\text{collection}} \approx 3.7\%\) for each detection arm. This factor is used to back-calculate the generated photon-pair rates from the measured coincidence counts.

The system is coupled with an Andor Shamrock 750 spectrometer body equipped with an iDus 420 CCD camera for spectral measurements. For the Raman measurements shown in  \textbf{Figure \ref{fig2}b and c}, the transmitted pump light was collected before the tunable long-pass (TLP) filter and directly coupled via a single-mode fiber into a spectrometer. No additional filtering was applied beyond the pump rejection by NF. The Raman spectra were recorded under the same focusing conditions as for the coincidence measurements. Transmission spectra (\textbf{Figure \ref{fig2}a}) were recorded using the transmission setup described in \cite{karaman_decoupling_nodate}. 

\textbf{Numerical Simulations:} Electric field distributions at the pump, signal, and idler wavelengths were calculated using the \textit{Wave Optics Module} of COMSOL Multiphysics. The metasurface and thin film geometries were modeled in 3D with the experimentally measured parameters, including a 100\,nm thick a-Si layer and periodic nanodisk arrays with diameters and pitches matching the fabricated samples. Wavelength-dependent refractive index data for a-Si were incorporated from ellipsometry measurements using interpolation functions. To emulate experimental conditions, \textit{Floquet periodic boundary conditions} were applied in the lateral directions, and perfectly matched layers (PMLs) were implemented along the propagation axis to absorb outgoing radiation.

The signal and idler mode fields were computed using the reciprocity approximation, where the outgoing modes are inferred from full-wave simulations with plane-wave illumination at the signal and idler wavelengths. This approach is valid in the undepleted-pump regime and allows us to approximate the quantum transition amplitude using classical fields. Simulations were performed for all three wavelengths involved in the SFWM process: pump (785\,nm), signal (depending on the detection band), and idler (depending on the detection band), using separate frequency-domain studies. The corresponding electric field distributions $\mathbf{E}_p(\mathbf{r})$, $\mathbf{E}_s^*(\mathbf{r})$, and $\mathbf{E}_i^*(\mathbf{r})$ were extracted and used to compute the nonlinear overlap integral:

\begin{equation}
\mathcal{O} = \int_V \chi^{(3)}(\mathbf{r}) \, \mathbf{E}_s^*(\mathbf{r}) \cdot \mathbf{E}_p(\mathbf{r}) \, \mathbf{E}_p(\mathbf{r}) \cdot \mathbf{E}_i^*(\mathbf{r}) \, d^3r,
\end{equation}

where $V$ is the nonlinear interaction volume. We assume that the dominant tensor element is $\chi^{(3)}_{xxxx}$, and co-polarized pump, signal, and idler fields are aligned along the same in-plane axis (e.g., $x$). This integral quantifies the spatial and polarization-mode matching between the pump, signal, and idler fields, and determines the photon-pair generation rate within the theoretical framework based on time-dependent perturbation theory. We also investigated the angular dependence of the metasurface response by simulating angle-resolved transmission spectra for both M1 and M2. These simulations confirmed that the Mie-type resonances are weakly dispersive within the collection cone defined by our high-NA (0.9) objective. Therefore, integrating signal and idler photons over this angular range does not significantly degrade the mode overlap or resonance alignment.

All field profiles were normalized to correspond to a fixed incident power, and the resulting $|\mathcal{O}|^2$ values were used to simulate the power-dependent coincidence rates. The effect of photo-thermal resonance shifts was also incorporated by re-running simulations at elevated temperatures, using temperature-dependent refractive index data obtained from separate ellipsometry measurements\cite{tagliabue_temperature_2025,karaman_decoupling_nodate}.

\section*{Data Availability Statement} \label{sec:data}
All the data supporting the findings of this study are presented in the Results section and Supplementary Information are available from the corresponding authors upon reasonable request.


\section*{Acknowledgements} \label{sec:acknowledgements}
G.T. acknowledge the support of the Swiss National Science Foundation (Starting Grant No. 211695). C.G. acknowledge the support of the Swiss National Science Foundation (SNSF project No. 214993). H.L. acknowledge the support of the National Natural Science Foundation of China (No. 22203042). The authors also acknowledge the support of the Center of MicroNanoTechnology (CMi) at EPFL.
\section*{Competing interests} \label{sec:comp}
The authors declare no competing financial interest

refs.bib file
%
\newpage

\bibliographystyle{unsrt}
\bibliography{main.bib} 
\end{document}